\documentclass[twoside,11pt]{article}
\usepackage{amssymb,amsmath}

\usepackage{graphicx, subfigure, wrapfig}
\usepackage{amsfonts}
\usepackage[english]{babel}

\RequirePackage{color}

\newtheorem{theorem}{Theorem}[section]

\newtheorem{remark}[theorem]{Remark}

\newtheorem{proposition}[theorem]{Proposition}
\newtheorem{corollary}[theorem]{Corollary}

\begin{document}

\title{On the Manev spatial isosceles three-body problem\\
Daniel Pa\c sca\footnote{Department of Mathematics and Informatics, University of Oradea, University Street 1, 410087 Oradea, Romania. Email: dpasca@uoradea.ro}  \,\,and\,\,Cristina Stoica\footnote{Department of Mathematics, Wilfrid Laurier University, Waterloo, Canada. Email: cstoica@wlu.ca}
}

%
%
%
\maketitle

\begin{abstract}
We study  the isosceles three-body problem with Manev interaction. Using a McGehee-type technique, we blow up the triple collision singularity into an invariant manifold, called the \textit{collision manifold},  pasted into the phase space for all energy levels. 
We  find that orbits tending to/ejecting from total collision  are present for a large set of angular momenta. We also discover that as the angular momentum is increased,  the collision manifold changes its topology. 
 \end{abstract}


\tableofcontents


\section{Introduction}
In 1930, the Bulgarian physicist Georgy Manev  proposed  a gravitational law of the form 
\begin{equation}
U(r)=-\frac{\mu}{r}-\left( \frac{3\mu^2}{2c^2}  \right)\frac{1}{r^2}
\end{equation}
where where $r$ is the distance between the bodies, $\mu$ the gravitational parameter, and $c$  the speed of light. He showed that by applying a general action-reaction principle to classical mechanics, one is  naturally led to the aforementioned law   \cite{Manev1, Manev2}. Provided the constants are chosen appropriately, the Manev  model can be used in calculations involving the perihelion advance of Mercury and  the other inner planets.

The $N$-body problem with Manev  interaction was brought into focus in the early 90's by  Diacu \cite{Diacu93}.  Due to its rich and interesting dynamics, it became  subject to many studies \cite{Diacu95, Diacu00, Szenkovits99, Stoica00, Diacu01, Santoprete02,  Puta05, Kyuldjiev07, Balsas09, Llibre12, Lemou12, Alberti15, Barrabes17}.  
For instance,  in contrast to its Newtonian counterpart, the Manev  problem displays binary collisions  for non-zero angular momenta: when approaching collision, two mass points spin infinitely many times around each other \cite{Diacu95, Diacu00}. In celestial mechanics community, this dynamical behaviour is known as a  \textit{black-hole}, somehow in analogy with the black-hole gravitational  effect \cite{Diacu95} in relativity.

  In the relative two-body problem, the Manev interaction delineates two distinct types of near-collision dynamics.  Let us consider the class of potentials of the form $-1/r-B/r^{\alpha}$, with $\alpha>1$, and $B>0$. {It can be shown  that for all $\alpha>1$ the double collision singularity may be blown up to a (smooth) torus, the equations of motions being  regularized (with a complete flow) \cite{Stoica00}. By the continuity with respect to initial data, the dynamics on the collision torus provides information on the behaviour of the near-collision flow.} 
For all $1<\alpha<2$, the   collision is possible only for zero angular momentum. The dynamics on the collision torus is similar to the Newtonian case, with a   gradient-like flow  matching two circles of  equilibria. Moreover, when   $\alpha= 2(1-1/n),$ $n\geq 2,$ $n \in \mathbb{N}$, the  flow is regularizable in the sense of  Levi-Civita \cite{Stoica02}. For $\alpha=2$,  i.e. in the Manev case, the   collision is possible   for angular momenta $C$ with  $|C|\leq b$, $b>0$ being some constant depending on masses;  
on the collision  torus,  the dynamics is trivial   displaying  two circles of degenerate equilibria \cite{Diacu00}.  For  $\alpha>2$ the collision torus is reached for all angular momenta. Its  flow is gradient-like, matching   two circles of equilibria as well, but it is not regularizable \cite{Stoica97}. 
An intuitive and physically reasonable explanation for the above is  that the Manev corrective term   $(-B/r^{2})$ adds  to the rotational inertial term $C^2/r^2$ (the latter being a consequence of the angular momentum conservation).

 We believe  that, similar to the two body problem with a potential of the form $-1/r-B/r^{\alpha}$ $\alpha>1$, $B>0$, in the $N$-body problem, the Manev $\alpha=2$ case marks the threshold between two distinct types of near-collision dynamics. The present work contributes to  clarification of this (see Section 6).

In this paper we investigate the dynamics  near total collision in a three-body problem with Manev binary interaction. {We note that an excellent exposition on the history together with most important results on the classical three body problem can be found online in the excellent \textit{scholarperia} article curated by \cite{Chenchiner07}.}

Considering  two of the masses equal, we study the dynamics   on  
 the invariant manifold of  isosceles configurations. Using  a McGehee technique similar to that in  \cite{Devaney80}, 
we  blow up the collision singularity  and replace it by an invariant \textit{collision manifold} pasted to the phase space for all energy levels. Due to the continuity of ODE solutions with respect to the initial data, the collision manifold, in spite of  fictitious,   provides information about orbits passing close to collision.
 Its (somplete) flow is rendered by the evolution of 3 variables, $v, \theta$ and $w$,  
describing the (fictitious) rate of change of the  size of the system, the shape of its configuration and the rate of change of the latter, respectively.

The Manev isosceles three-body problem, and in particular the near-collision dynamics,  was studied 
 by Diacu  \cite{Diacu93}, but only for zero total  angular momentum. In this case, the three bodies were confined to a fixed plane, with the middle body oscillating above and below  the line joining the other two. 
 One of the open problems   stated in Diacu's paper concerns the existence of \textit{non-zero  angular momenta} orbits ejecting/tending asymptotically to triple collision. Are such orbits possible?  Here we give a positive answer to this question, showing that these orbits exist  for a large set (in fact, an interval) of non-zero momenta.

 We also detect an interesting feature of the Manev three-body problem: as the size $C$ of the total angular momentum  increases from zero,  the  collision  manifold  changes its topology  from a sphere with 4 points removed, as in the Newtonian \cite{Shibayama09} and Schwarzschild \cite{Arredondo14} cases, to the union of a sphere with two lines,  to the union a point with two lines, and finally to two lines. To our knowledge, this phenomenon was not observed anywhere else.  The lines that persist for all momenta correspond to (fictitious) double collisions.

On the collision manifold,   the double collisions lines are filled with  equilibria for all $C$. For low momenta,  we find six  more equilibria, similar to the Newtonian case \cite{Shibayama09}.  These points correspond to two distinct total collision limit configurations: one linear (with 
one of the body fixed on the midpoint between the other two) and one  spatial (modulo a reflection symmetry), with the ratio of the triangle sides depending on the bodies' masses. 
As $C$ is increased, the spatial limit configurations disappear. For  high $C$, the linear limit configurations disappear as well and  triple collision  is reached (asymptotically) only by solutions with    limit configurations in double collisions.

The flow on ${\cal C}$ is constant in the $v$ coordinate:  for low $C$,  the orbits connect the double collision manifolds, whereas for higher $C$, when ${\cal C}$ is diffeomorphic to the union of a sphere with the double collision lines, all orbits are either periodic or equilibria. 
 For a fixed negative level of the total energy $h<0$, we  prove that  these periodic orbits are not  attractors for the global flow; {this implies that \textit{in this case},  there are no triple collision orbits trapped asymptotically in a near-collision quasi-periodic behaviour. Specifically, there are no orbits for which the outer masses  rotate (with increasing spin) about the centre of mass and the middle mass oscillates up and down, with the triple collision being attained asymptotically in time.} We  also prove that homographic motions, that is motions for with self-similar configurations, have linear configurations only. Finally, we observe that for strictly positive total energy $h>0$, all orbits are unbounded.

  \smallskip
  
 The paper is organized as follows: in Section 2 we introduce the isosceles Manev three-body problem and reduce the dynamics to a two degrees of freedom using the angular momentum conservation. In Section 3 we regularize the equations of motion.  In Section 4 we define the collision manifold, classify its topology and  investigate the associated  dynamics.  In Section 5 we discuss the flow  near-by the collision manifold, study equilibria and homographic motions, and prove some statements on  the global flow. {We end in Section 6  by presenting a brief comparison of the near-collision dynamics as in the potential formula $-1/r-B/r^{\alpha}$, $B>0$, the parameter  $\alpha$ increases  from $\alpha=1$ to $\alpha=3$.}

\section{Dynamics}
\begin{figure}[h!]
\centerline
{\includegraphics[scale=0.67]{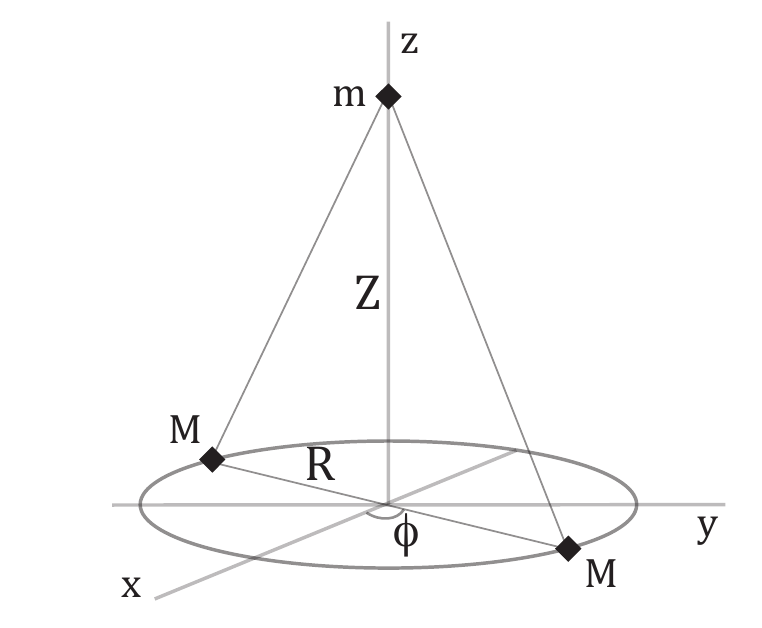}}
{\caption{The spatial isosceles three-body problem
\label{Figure}
}}
\end{figure}

 In cylindrical coordinates $(R, \phi, Z, p_R, p_{\phi}, p_Z)$ (see Figure \ref{Figure}) the Hamiltonian is
\begin{align*}
  H(R, \phi, Z, P_{R}, P_{\phi}, P_Z)= \frac{1}{M} &\left( P_R^2+ \frac{P_{\phi}^2}{R^2}\right) \\
  &+ \frac{2M+m} {4Mm} P_{Z}^2 +
  U(R, Z ),
  \end{align*}
with a Manev-type potential given by
 \begin{align}
U(R, Z)&= -\frac{GM^2}{R} \left(1+\frac{\gamma_0}{R}\right) 
\nonumber \\
&-\frac{4GMm}{\sqrt{R^2+4Z^2}}
\left( 1+\frac{4\gamma}{\sqrt{R^2+4Z^2} }\right),
\label{P_cyl}
\end{align}
where $\gamma_0, \gamma>0$ and $\gamma_0\neq\gamma_0$. For reason to be discussed later, we assume that 
\begin{equation}\label{pf3}
16\gamma >\gamma_0\,.
\end{equation}
Using the angular momentum conservation
$P_{\phi}(t)=const.=:C\,$
we reduced the dynamics to  a two degree of freedom Hamiltonian system  determined by
 \begin{align}
 &H_{\text{red}}(R,  Z, P_{R},  P_Z; C)    \label{H_red}
\\
&=\frac{1}{2}(p_R  \quad  p_{z}) \left(
\begin{array}{cc}
\frac{2}{M} &0\\
0&\frac{2M+m} {2Mm}
\end{array}
\right)
\left(
 \begin{array}{c}
P_R\\
P_Z
\end{array}
 \right) 
+
  U_{\text{eff}}(R, Z; C) 
  \nonumber
  \end{align}
where  $U_{\text{eff}}(R, Z; C)$ the effective (or amended) potential
   \begin{equation}\label{U_eff}
   U_{\text{eff}}(R, Z; C):=\frac{C^2}{MR^2}  + U(R,Z).
      \end{equation}
and $C\in\mathbb{R}$ is a parameter.
The equations of motion are
   \begin{align*}
&\dot R = \frac{2P_R}{M}\,, \\
&\dot Z = \frac{2M+m}{2Mm}P_Z\,,  \\
&\dot P_R =  \frac{2C^2}{MR^3} -\frac{\partial U(R,Z;C)}{\partial R}\\
&\dot P_Z=  -\frac{\partial U(R,Z;C)}{\partial Z}\,.
   \end{align*}
Since the Hamiltonian is time-independent, along any solution the energy is conserved:
   \begin{equation}\label{H_red_eff}
H_{\text{red}}\left(R(t),  Z(t), P_{R}(t),  P_z(t); C\right)=const.=h.
\end{equation}

\section{The regularized dynamics}\label{CM.sect}
We now regularize  the equations of motion. For this, 
we follow closely the McGehee technique as used in the  Newtonian isosceles  problem by Devaney \cite{Devaney80}. Denoting
\[{\bf x}:=
\left(
\begin{array}{c}
R\\
Z
\end{array}
\right), \, {\bf p}:=
 \left(
\begin{array}{c}
p_R\\
p_Z
\end{array}
\right),\,\mathbb{K}=
\left(
\begin{array}{cc}
\frac{M}{2} &0\\
0&\frac{2Mm}{2M+m}
\end{array}
\right)\,,
\]
we introduce the  coordinates $(r,v, {\bf {s}}, {\bf u} )$  defined by
\begin{align}\label{sys0}
r&=\sqrt{{\bf x}^{t}\mathbb{K}\,{\bf x}},\quad \quad \quad
v=r({\bf s}\cdot {\bf p}),\\
{\bf s}&=\frac{{\bf x}}{r},\quad \quad \quad  \quad \quad \,\,\,
{\bf u}=r (\mathbb{K}^{-1}{\bf p}-({\bf s}\cdot {\bf p}) {\bf s}).\nonumber
\end{align}
%
%
%
Notice that $r=0$ corresponds to $R=Z=0$, i.e., to the triple collision of the bodies. The coordinate $v$ describes the rate of change of the size of the system as given by $r$, whereas  the vector ${\bf s}$ describes $R$ and $Z$ separately.  One may verify that in the new coordinates we have
that ${\bf s}^t\mathbb{K}\,{\bf s}=1$ and ${\bf s}^t\mathbb{K}\,{\bf u}=0$. The equations of motion are
\begin{align*}
&\dot{r}=  r^{-1} v,\\
&\dot{v}= r^{-2} v^2 + r^{-2} {\bf u}^t \mathbb{K} {\bf u} +
 r^{-2} \frac{2C^2}{Ms_{1}^2} -  \left( \frac{ V({\bf s})}{r} + \frac{2 W({\bf s})}{r^2} \right),\\
&\dot {\bf s}= r^{-2} {\bf u},\\
&\dot {\bf u}=  \left[- r^{-2} {\bf u}^t \mathbb{K} {\bf u} - r^{-2}
\frac{2C^2}{M s_1^2} + r^{-2} \left( \frac{V({\bf s})}{r} + \frac{2 W({\bf s})}{r^2} \right) \right] {\bf s}\\
\,\\
&+ r^{-1} \left(
                            \begin{array}{c}
                              \dfrac{2}{M}\dfrac{\partial V}{\partial s_1}  \\
                                         \dfrac{2M+m}{2Mm}\dfrac{\partial V}{\partial s_2}  \\
                                       \end{array}
                                     \right)+
                                   r^{-2} \left(
                                     \begin{array}{c}
                                     \dfrac{\partial }{\partial s_1} \left(  - \frac{2 C^2}{M^2s_1^2} \right)\\
                                     0
                                       \end{array}
                                     \right)+\\
& \hspace{4cm}r^{-2} \left(
                                       \begin{array}{c}
                                         \dfrac{2}{M}\dfrac{\partial W}{\partial s_1}  \\
                                         \dfrac{2M+m}{2Mm}\dfrac{\partial W}{\partial s_2}  \\
                                       \end{array}
                                     \right),
\end{align*}
with
\[V({\bf s}) = \frac{G M^2}{s_1} + \frac{4G  Mm}{\left( s_1^2+4s_2^2 \right)^{\frac{1}{2}}} \quad \,\,\text{and} \]
\[W({\bf s}) = \frac{G M^2 \gamma_0}{s_1^2} + \frac{8G  M m \gamma}{\left( s_1^2+4s_2^2 \right)}.\]
We further introduce the change of coordinates
\[{ \bf s}=\sqrt{(\mathbb{K}^{-1})}(\cos\theta,\sin\theta)^t,\,{\bf u}=u\sqrt{(\mathbb{K}^{-1})}(-\sin\theta,\cos\theta)^t\]
where $\displaystyle{-\dfrac{\pi}{2}<\theta<\dfrac{\pi}{2}}$ so that the boundaries $\displaystyle{\theta=\pm \frac{\pi}{2}}$ correspond in the original coordinates to $R=0,$ that is, to  double collisions of the masses $M.$ 
More precisely, at $\theta=\pi/2$ we have $R=0$ and $z>0$, whereas at $\theta=-\pi/2$, $R=0$ and $z<0$. Also, the  $\theta$ varies, the ratio between $R$ and $Z$ varies as well; a direct calculation also shows that
\begin{equation}
Z \cos \theta = \frac{\sqrt{\mu}}{2}R \sin \theta\,.
\label{ratio}
\end{equation}
Thus, for instance,  $Z=0$ at $\theta=0$, and $R=0$ at $\pm \theta=\pi/2$. One may also verify that  that  ${\bf u}^t T {\bf u}= u^2$ and
$\displaystyle{
\dot {\bf u}=({\dot u}/{u})   {\bf u}  - u \,\dot \theta \,{\bf s}.}
$
%
Denoting
 \begin{equation}\label{mu}
 \mu:=\dfrac{2M+m}{m}\,
 \end{equation}
and applying the time re-parametrization  $dt=r^{2}d\tau$, we obtain the system
\begin{align}
r^{\prime}&=rv,\label{sys_MP-1} \\
v^{\prime}&=v^2+u^{2}+\frac{C^2}{\cos^2\theta} -r  V(\theta) - 2 W(\theta), \label{sys_MP-2} \\
\theta^{\prime}&=u , \label{sys_MP-3}\\
u^{\prime}&=\ - C^2\frac{\sin\theta}{\cos^3\theta} + r\,\frac{\partial V(\theta)}{\partial \theta} +\frac{\partial W(\theta)}{\partial \theta}, \label{sys_MP-4}
\end{align}
where
\begin{align}
 \label{V_de_theta}
V(\theta)&=GM\left(\frac{M}{2}\right)^{\frac{1}{2}}\left(\frac{M}{\cos\theta}+\frac{4m }{(\cos^2\theta+\mu\sin^2\theta)^{\frac{1}{2}}}\right),\\
W(\theta)&=G M\left(\frac{M}{2}\right)\left(\frac{M \gamma_0}{\cos^2\theta}+\frac{8m \gamma}{(\cos^2\theta+\mu\sin^2\theta)}\right).  
\label{W_de_theta}
\end{align}
In the new coordinates the energy integral is given by
\begin{equation}\label{e_r}
hr^2= \frac{1}{2} \left(  u^2+v^2 \right)  - r V(\theta) - W(\theta)\,.
\end{equation}

\subsection{Potential functions $V(\theta)$ and $W(\theta)$}\label{PF.subsection}

First we notice that $V(\theta)$ and $W(\theta)$ are positive on their domain $\theta \in (-\pi/2, \pi/2)$. A direct calculation shows that, $V(\theta)$ has three critical points at $\theta_0 = 0$ and $\theta = \pm\theta_v$, where
\begin{equation}\label{pf2}
\cos \theta_v = \sqrt{\frac{\mu}{\mu+3}}\,.
\end{equation}
\begin{figure}[h]
\centerline
{\includegraphics[scale=0.3]{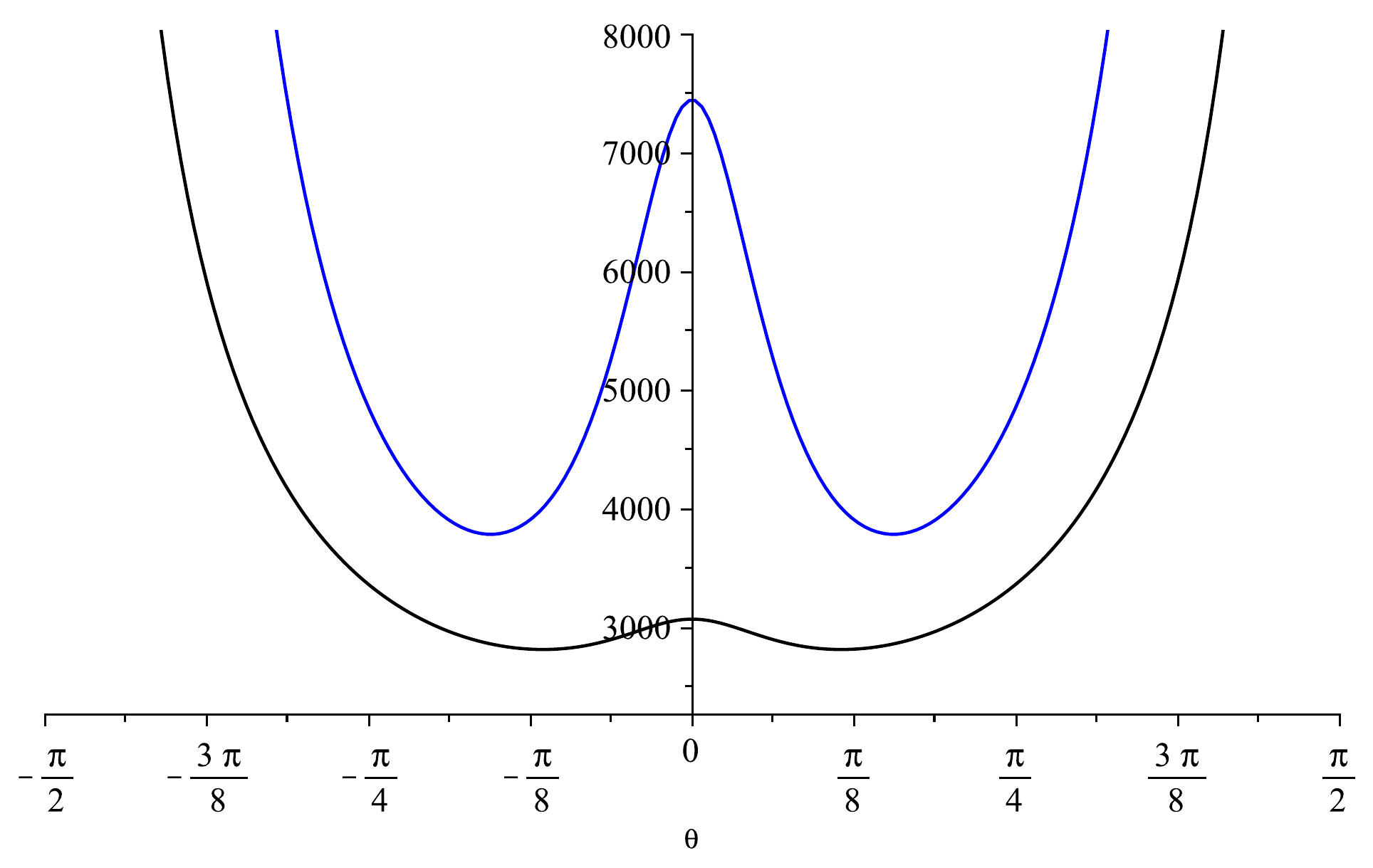}}
{\caption{The shape and the intersection of the functions $V(\theta)$  (bottom) and $W(\theta)$ (top).
(The figure is generated for $M=10$,  $m=1$, $\gamma_0=1$ and $\gamma=3$.) 
\label{V-W-plot}
}}
\end{figure}
Similarly,  provided the conditions \eqref{pf3} is satisfied, $W(\theta)$
displays three critical points at $\theta_0 = 0$ and $\theta=\theta_w$, where
\begin{equation}\label{pf4}
\cos \theta_w =
\sqrt{\frac{\mu}{\mu+4 \sqrt{\frac{\gamma}{\gamma_0}} -1 }}\,.
\end{equation}
We leave for future work the case  when the parameters $\gamma_0$ and $\gamma$  do not obey \eqref{pf3} (that is when $\gamma_0\geq 16\gamma$). 
It is immediate that the nonzero critical points of $V(\theta)$ and $W(\theta)$ coincide only if $\gamma=\gamma_0,$ case already excluded in our model; see equation  \eqref{P_cyl}.

\subsection{Regularized Equations of Motion}
\label{TCM.subsection}

\begin{figure}[h]
\centerline
{\includegraphics[scale=0.23]{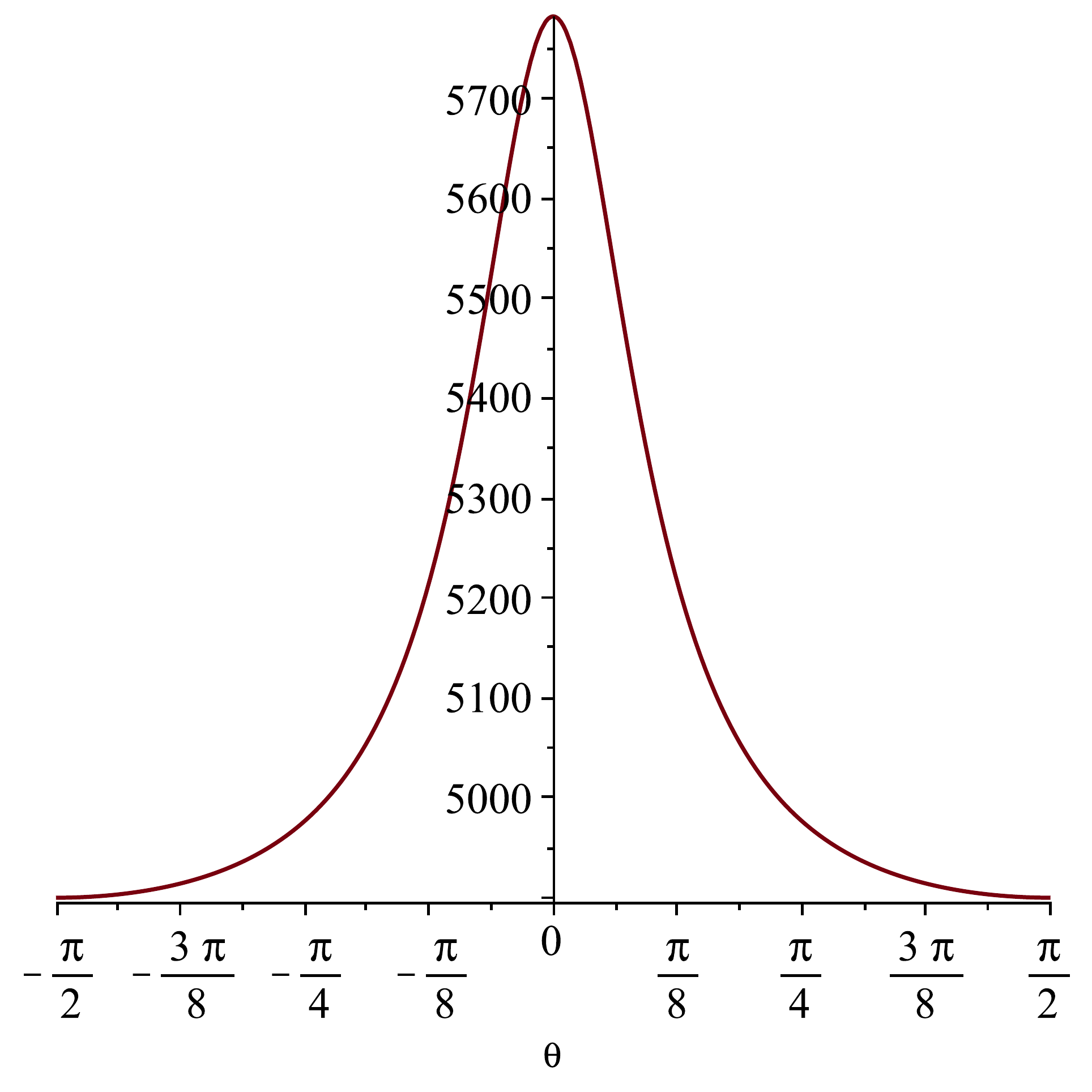}}
{\caption{The function $U(\theta).$
\label{U_theta}
}}
\end{figure}

In the system \eqref{sys_MP-1}-\eqref{sys_MP-3} and the energy integral \eqref{e_r}
we make the substitutions
\begin{equation}\label{subs}
U(\theta) = W(\theta)\cos^2 \theta, \,\,\, w = \frac{\cos^2 \theta}{\sqrt{U(\theta)}}u,
\end{equation}
and introduce a new time parametrization given by $\frac{d \tau}{d \sigma} = \frac{\cos^2 \theta}{\sqrt{U(\theta)}}$ to obtain
\begin{align}\label{sys_MP1}
r^{\prime}&=\frac{\cos^2 \theta}{\sqrt{U(\theta)}}rv,\nonumber \\
v^{\prime}&= \Big( v^2 + \frac{U(\theta)}{\cos^4 \theta} w^2+ \frac{C^2}{\cos^2\theta} -r V(\theta) - 2 \frac{U(\theta)}{\cos^2 \theta}
\Big) \frac{\cos^2 \theta}{\sqrt{U(\theta)}}, \\
\theta^{\prime}&=w ,\nonumber\\
w^{\prime}&= - C^2\frac{\sin2\theta}{2 U(\theta)} + r \cos^4 \theta\frac{V'(\theta)}{U(\theta)}+\cos^2 \theta
\frac{U'(\theta)}{U(\theta)} +\sin 2\theta,\nonumber
\end{align}
and
\begin{align}
2hr^2 \cos^4 \theta =  w^2 &U(\theta)+v^2 \cos^4 \theta  + C^2 \cos^2\theta  \nonumber\\
&- 2r \cos^4 \, \theta V(\theta) - 2 \cos^2 \, \theta U(\theta)\,.
\label{e_r_MP1}
\end{align}
Notice that $U(\theta)$ is smooth and $U(\theta)>0$ for all $\theta\in (-\pi/2, \pi/2)$; see its sketch in Figure \ref{U_theta}.
Finally, using the energy relation, we substitute the term containing the angular
momentum $C$ into the $v^{\prime}$ equation and obtain
\begin{align}
r^{\prime}&=\frac{\cos^2 \theta}{\sqrt{U(\theta)}}rv, \label{sys_MP2} \\
v^{\prime}&= r(2hr + V(\theta)) \frac{\cos^2 \theta}{\sqrt{U(\theta)}}, \label{sys_MP2-1}\\
\theta^{\prime}&=w ,  \label{sys_MP2-11}\\
w^{\prime}&= \frac{\cos\theta}{U(\theta)}\left( r V'(\theta)  \cos^3 \theta 
+ U'(\theta)\cos \theta \right.  \nonumber \\
&\,\,\, \quad \quad \quad \quad \quad \quad \quad \quad \quad \left. - \left(C^2 -2{U(\theta)}\right)\sin\theta 
 \right). \label{sys_MP2-2}
\end{align}

\section{The Triple Collision Manifold}

The vector field \eqref{sys_MP2}-\eqref{sys_MP2-2} is analytic on \newline $\displaystyle{[0,\infty)\times \mathbb{R}\times \left(-\frac{\pi}{2}, \frac{\pi}{2} \right)\times \mathbb{R}}$,
and thus the flow is well defined everywhere on its domain, including the points corresponding to triple collision
$(r = 0)$.  
 The restriction of the energy relation (\ref{e_r_MP1}) to $r = 0$ 
\begin{align}
{\cal C} :=& \left\{ (v,\theta,w) \in  \mathbb{R}\times \left[-\frac{\pi}{2}, \frac{\pi}{2} \right]\times \mathbb{R}\, \Big|
 \right. \nonumber \\
&\left.
 w^2 +v^2 \frac{\cos^4 \, \theta}{U(\theta)}  + \left(C^2 - 2U(\theta)\right)\frac{\cos^2 \theta}{U(\theta)} =0 \right\},
\label{ColMan}
\end{align}
is  a (fictitious) invariant set, called the {\it triple collision manifold}, pasted into
the phase space for any level of energy.   
By continuity with
respect to the initial data, the flow on  the smooth subsets of  ${\cal C}$  provides information about the orbits that pass close to collision.

\subsection{Topology}\label{Col_man}

Let is denote by $U_m$  the minimum and maximum values of $U(\theta)$ (see Figure \ref{U_theta}). We calculate
\begin{equation}
U_m=U\left(\pm \frac{\pi}{2}\right) = \frac{GM^3 \gamma_0}{2}\,. 
\label{U_min}
\end{equation}
We also observe that the maximum value of $U(\theta)$ occurs at $\theta=0$ and it is given by
\begin{equation}
U(0) = \frac{GM^2}{2}(M\gamma_0 + 8m\gamma)\,.
\label{U_Max}
\end{equation}
The collision manifold is non-void if $\displaystyle{\left(C^2-2U(\theta) \right)\leq 0}.$ Considering  the graph of $2U(\theta)$ and  the sign of $\displaystyle{\left(C^2-2U(\theta) \right)}$ as $C^2$ is increasing from zero, we distinguish the following cases:
\begin{enumerate}

\item If $\displaystyle{0 \leq  |C|<\sqrt{2 U_m}}$ the collision manifold ${\cal C}$  is homeomorphic to a sphere with 4 points removed; see Figure  \ref{CM_small}.  ${\cal C}$ is a smooth manifold everywhere, except at the (fictitious) double collision boundaries 
\begin{equation}
{\cal B}_{l,r}:=\{(v, \theta, w)\,|\, v=v_0 \in \mathbb{R}\,, \theta=\pm \frac{\pi}{2}\, w=0\}\,.
\end{equation}
\begin{figure}[h]
\centerline
{\includegraphics[scale=0.37]{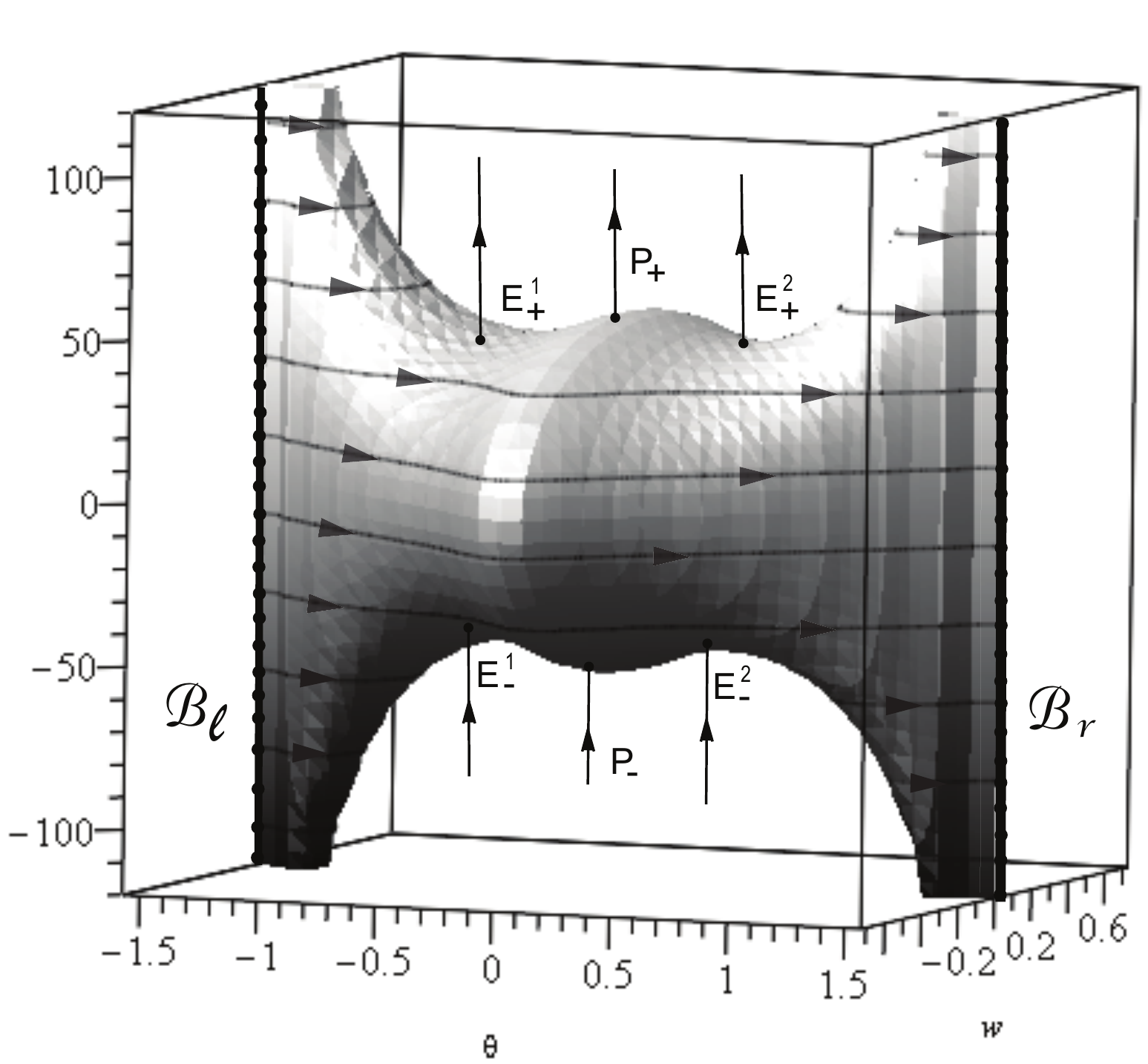}}
{\caption{The collision manifold ${\cal C}$ for angular momenta $\displaystyle{0\leq |C|\leq \sqrt{2 U_m}}.$
\label{CM_small}
}}
\end{figure}

\item If $\displaystyle{|C|\in\left(\sqrt{2 U_m}\,, \sqrt{2U(0)}  \right)}$ then ${\cal C}$ consists in  the union of  a sphere with the lines ${\cal B}_{l,r}$; 
see Figure \ref{CM_C_in_Um_UM}.

\smallskip

\item If $\displaystyle{|C|= \sqrt{2U(0)}}$ then ${\cal C}$  is the union of    one point,  the origin, with ${\cal B}_{l,r}$.

\smallskip

\item If $\displaystyle{|C|> \sqrt{2U(0)}}$ then ${\cal C}$ consists of the lines  ${\cal B}_{l,r}$.

\begin{figure}[h]
\centerline
{\includegraphics[scale=0.51]{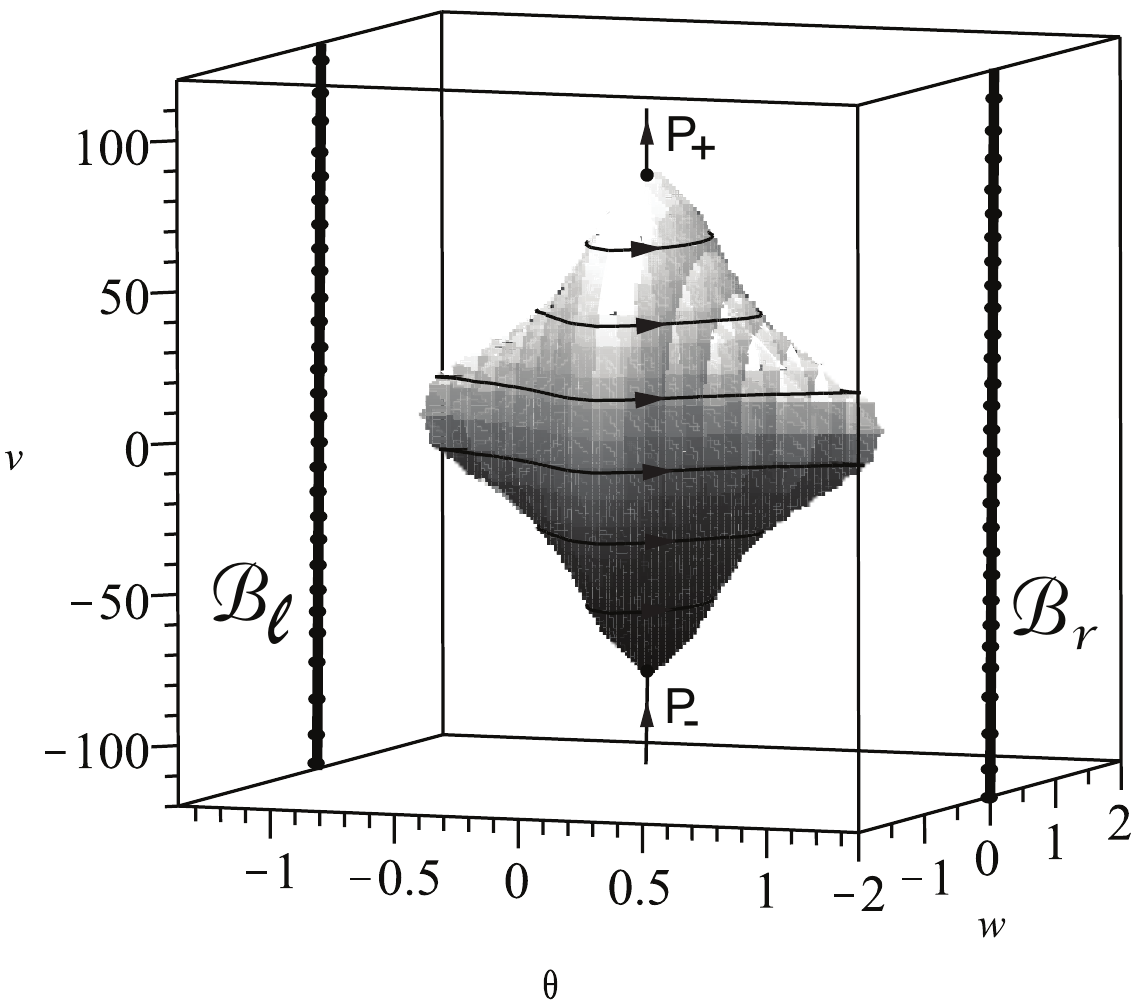}}
{\caption{The collision manifold ${\cal C}$ for angular momenta $\displaystyle{|C|\in\left(  \sqrt{2 U_m}\,,\sqrt{2U(0)}  \right)}.$ The compact part  ${\cal C} \setminus {\cal B}_{l,r}$ of the collision manifold shrinks as the total angular momentum $|C|$ is increasing,  and it disappears for $|C|> \sqrt{2U(0)}.$
\label{CM_C_in_Um_UM}
}}
\end{figure}

\end{enumerate}

Thus we have proved:

\smallskip
\begin{proposition} 
 As the momentum $|C|$ is increased, the triple collision manifold  changes its topology, from a sphere with 4 points removed, to the union of  a sphere with two lines, to the union of  a point with two lines and finally, to two lines. 
\end{proposition}

 \smallskip

\subsection{Dynamics on the collision manifold}\label{CME}

The vector field on the
collision manifold  is obtained by setting $r = 0$ in system (\ref{sys_MP2}) and it is given by
\begin{align}\label{s_1}
v^{\prime}&= 0, \\
\theta^{\prime}&=w ,\label{s_2}\\
w^{\prime}&=\frac{\cos\theta}{U(\theta)}\left(  U'(\theta)\cos \theta - \left(  C^2 -2{U(\theta)}\right)\sin\theta \right)\,.
\label{s_3}
\end{align}
%
%
It is immediate  that $v(t)$ is constant along the   orbits, the flow being degenerate in this direction. The restriction of collision manifold ${\cal C}$ to a  level $v_0=:v(t)$ is 
\begin{align}
{\cal V}_{v_0}& := \left\{ (\theta,w) \in  \left(-\frac{\pi}{2}, \frac{\pi}{2} \right)\times \mathbb{R}\, \Big| 
\right. \nonumber \\ 
&\left.
w^2 +v_0^2 \frac{\cos^4 \, \theta}{U(\theta)}  + \left(C^2 - 2U(\theta)\right)\frac{\cos^2 \theta}{U(\theta)} =0 \right\}\,.
\label{ColMan-1}
\end{align}
When connected to ${\cal C}$, the double collision lines ${\cal B}$
consist in  degenerate equilibria.  All orbits are horizontal. 

\smallskip
\noindent
For all momentum values for ${\cal C}$ exists, that is for 
\begin{equation}
|C|\leq \sqrt{2U(0)}=\sqrt{ GM^3\gamma_0 +  8GM^2 \gamma}
\end{equation}
we have   two equilibria located at
\begin{equation}
P_{\pm}:=(\pm \sqrt{2U(0)-C^2}, 0, 0)\,.
\label{Ps}
\end{equation}
\smallskip
\noindent
For  momenta 
\begin{equation}
|C|= \sqrt{2U_m}=  \sqrt{GM^3\gamma_0}
\end{equation}
the equilibria $P_{\pm}$  coalesce.

\smallskip
\noindent
For lower momenta 
\begin{equation}
|C|\leq \sqrt{2U_m}=  \sqrt{GM^3\gamma_0}
\end{equation}
we also have four more equilibria located at
\begin{equation}
E^{1}_{\pm}=(\pm v_0, -\theta_0, 0)\,\quad\text{and} \quad E^{2}_{\pm}=(\pm v_0, \theta_0, 0)
\end{equation}
where
\begin{equation}
v_0 = \frac{1}{\mu} \left[\sqrt{8GM^2m\gamma} + \sqrt{\frac{2M}{m} \left(  GM^3\gamma_0 -C^2 \right)} \right]
\label{v_0}
\end{equation}
and $\theta_0 \in (0, \pi/2)$ so that 
\begin{equation}
\tan^2 \theta_0 = \frac{1}{\mu} \left(\sqrt{\frac{16GM^3\gamma}{GM^3\gamma_0-C^2}}-1 \right).
\label{theta_0}
\end{equation}

\smallskip
\noindent
Consequently,   for $|C|\leq  \sqrt{GM^3\gamma_0}$
we have the following type of orbits (see Figure \ref{CM_small}):

\smallskip
- homoclinic connections joining a double collision equilibrium;

\smallskip
- heteroclinic connections joining  a double collision equilibrium to one of the $``E"$points;

\smallskip
- homoclinic connections between two $``E"$ points;

\smallskip
- heteroclinic connections joining two double collision equilibria\,.

\smallskip
\noindent
 On the edges ${\cal B}_{l,r}$  the system   \eqref{s_1}-\eqref{s_3} may lose uniqueness of solutions. The  double collisions are not regularizable (and thus they cannot be equivalent to elastic bounces, as in the Newtonian case), as it is known from  the   \cite{Diacu00, Stoica00}.

\smallskip
\noindent
For $\sqrt{2U_m} < |C|<  \sqrt{2U(0)} $, that is for
\[\sqrt{GM^3\gamma_0} < |C|< \sqrt{GM^3 \gamma_0 + 8GM^2m \gamma}\]
the flow wraps around ${\cal C}$ (see Figure \ref{CM_C_in_Um_UM}).

\section{The Near-Collision Flow}

\subsection{Equilibria and their stability}

We now discuss the equilibria on the collision manifold as embedded in  the full $(r,v, \theta, w)$ regularized phase-space, and calculate their stability.  We have

\smallskip
\noindent
- for all momenta $|C|\leq \sqrt{2U(0)}$, we find a pair equilibria on  ${\cal C}$  at
\begin{equation}
P_{\pm}:=(0, \pm \sqrt{2U(0)-C^2}, 0, 0)\,.
\end{equation}

\smallskip
\noindent
- for  momenta such $|C|\leq \sqrt{2U_m}$ the flow  also displays four fixed points 
\begin{equation}
E^{1}_{\pm}=(0, \pm v_0, -\theta_0, 0)\quad \quad E^{2}_{\pm}=(0, \pm v_0, \theta_0, 0)
\end{equation}
with $v_0$ and $\theta_0$ given by \eqref{v_0} and \eqref{theta_0}, respectively.
Also,  we find an infinite number of equilibria on the edges ${\cal B}_{l.r}$.

\smallskip
To determine the stability of $P_{\pm}$ we start by writing the energy relation (\ref{e_r_MP1}) as a level set:
\begin{equation}\label{l_e}
{\cal E}:=\{(r,v,\theta,w)\,|\, F(r,v,\theta,w)=0\}
\end{equation}
where 
%
\begin{align}
F(r,&v,\theta,w):=2hr^2 \cos^4 \theta 
-  w^2 U(\theta) - v^2 \cos^4 \theta  \nonumber \\
& - C^2 \cos^2\theta  + 2r V(\theta) \cos^4 \, \theta  + 2 U(\theta)\cos^2 \, \theta \,.
\end{align}
Next we calculate the spectrum of the linearization of system \eqref{sys_MP2} at an equilibrium and then we restrict it  to the tangent space of the collision manifold ${\cal E}$. We denote by $J$ the linearization of \ref{sys_MP2} and  $\bar{J}$ its restriction to a tangent space.

\medskip
At $P_{\pm}=(0, \pm \sqrt{2U(0)-C^2}, 0, 0)$ we find
\begin{equation}\label{J}
J=\left(
\begin{array}{cccc}
\pm \sqrt{2-\frac{C^2}{U(0)}} & 0 & 0 & 0 \\ \\
\frac{V(0)}{\sqrt{U(0)}} & 0 & 0 & 0 \\ \\
0 & 0 & 0 & 1 \\ \\
0 & 0 & 2 + \frac{U''(0)-C^2}{U(0)}  & 0
\end{array}
\right).
\end{equation}
The tangent space to ${\cal E}$ at an equilibrium point $P_{\pm}=(0, \pm \sqrt{2U(0)-C^2}, 0, 0)$ is
\begin{align}
&T_{P_{\pm}} {\cal E} = \{(\rho_1, \rho_2, \rho_3, \rho_4)\,|\, \nabla F|_{P_{\pm}} \,\cdot \, (\rho_1, \rho_2, \rho_3, \rho_4) =0 \} \nonumber \\
& \quad \,= \{(\rho_1, \rho_2, \rho_3, \rho_4)\,|\, V(0) \rho_1 \pm \sqrt{2U(0) - C^2} \rho_2 =0 \}\, . \nonumber
\end{align}

\noindent
For angular momenta 
 \[|C| < \sqrt{2U(0)}\,= G M^2(M\gamma_0 + 8m\gamma)/2,\] 
a basis for $T_{P_{\pm}} {\cal E}$ is given by 
\[\xi_1 = (\pm \sqrt{2U(0)-C^2}, -V(0), 0, 0),\]
$\xi_3 = (0, 0, 1, 0)$ and $\xi_4 = (0, 0, 0, 1)$ and 
a representative of $\bar{J}$ in this basis is
\begin{equation}\label{J_res_rep_2}
\left(
\begin{array}{ccc}
\pm \sqrt{2-\frac{C^2}{U(0)}}&0&0\\
0& 0 & 1 \\
0& 2 + \frac{U''(0)-C^2}{U(0)} & 0
\end{array}
\right).
\end{equation}
The eigenvalues of $\bar J$ are given by
 \[
 \lambda_1 = \pm \sqrt{2-\frac{C^2}{U(0)}} =  \pm
  \sqrt{\frac{GM^3\gamma_0- C^2}{GM^3\gamma_0}}
  \in \mathbb{R}
 \]
 and 
\begin{equation}
\lambda_{2,3} =\pm i \sqrt{\frac{(-2) \left(GM^3(\gamma_0 - 16\gamma) -C^2 \right)}{GM^2(M\gamma_0 + 8m\gamma)}}
\end{equation}
where the quantity under square root is positive given that condition (\ref{pf3}) is satisfied.

\smallskip
\noindent
If  $|C| =\pm \sqrt{2U(0)}$,   the collision manifold collapses to a point, the origin $O$, which is also an equilibrium. We have
$T_{O} {\cal E} = \{(\rho_1, \rho_2, \rho_3, \rho_4)\,|\, \rho_1 =0 \}$. The linear part of the vector field (\ref{sys_MP2}) restricted to the
tangent space is given by
\begin{equation}\label{J_res}
\bar{J}=\left(
\begin{array}{cccc}
0 & 0 & 0 & 0 \\ \\
0 & 0 & 0 & 0 \\ \\
0 & 0 & 0 & 1 \\ \\
0 & 0 & \frac{U''(0)}{U(0)}  & 0
\end{array}
\right),
\end{equation}
and so a basis for $T_{\pm P}{\cal E}$ is given by $\xi_2 = (0, 1, 0, 0)$, $\xi_3 = (0, 0, 1, 0)$ and $\xi_4 = (0, 0, 0, 1).$
A representative of $\bar{J}$ in this basis is
\begin{equation}\label{J_res_rep_1}
\left(
\begin{array}{ccc}
0&0&0\\
0& 0 & 1 \\
0& \frac{U''(0)}{U(0)}  & 0
\end{array}
\right).
\end{equation}
The eigenvalues are given by  $\lambda_1=0$ and 
\[
\lambda_{2,3} = \pm 4 i \sqrt{m\gamma \mu /(M\gamma_0 + 8m \gamma)}\,.
\]

\smallskip
\noindent
Now we study the behaviour near the points  $E^{1,2}_{\pm}.$ We calculate 
the Jacobian matrix of system \eqref{sys_MP2} evaluated at this points and find:
\begin{equation}\label{J_1}
J=\left(
\begin{array}{cccc}
\frac{\pm v_0 \cos^2 \theta_0}{\sqrt{U(\theta_0)}} & 0 & 0 & 0 \\ \\
\frac{V(\theta_0) \cos^2 \theta_0}{\sqrt{U(\theta_0)}}  & 0 & 0 & 0 \\ \\
0 & 0 & 0 & 1 \\ \\
\frac{V'(\theta_0) \cos^4 \theta_0}{U(\theta_0)} & 0 & a & 0
\end{array}
\right)
\end{equation}
where 
\begin{align}
a=&\frac{
16Mm^2(2M+m)\gamma \sin^2 \theta_0 \cos^4 \theta_0
}{
\left( M\cos^2\theta_0 -M-\frac{m}{2}  \right)^2
}\nonumber \\
&\,   \frac{1}{\left[
\left(M^2\gamma_0-4m^2 \gamma
\right)\cos^2 \theta_0
-M\left(M+\frac{m}{2}  \right)\gamma_0
\right]}\,.
\end{align}
The sign of the term $a$ is decided by the sign of the expression
\begin{equation}
T:=\left(M^2\gamma_0-4m^2 \gamma
\right)\cos^2 \theta_0
-M\left(M+\frac{m}{2}  \right)\gamma_0\,.
\label{expresie}
\end{equation}
For this we calculate $\cos^2\theta_0=1/(1+\tan^2\theta_0)$ using \eqref{theta_0} that we then substitute into \eqref{expresie}. We obtain
\begin{equation}
T= - \frac{m(2M+m)\Big(8m\gamma + M\gamma_0 \sqrt{\frac{16GM^3 \gamma}{GM^3\gamma_0-C^2}}\Big)}{2\Big(2M + m
\sqrt{\frac{16GM^3 \gamma}{GM^3\gamma_0-C^2}}\Big)}.
\end{equation}
Thus the sign of $a$ is negative. The tangent space to the energy level manifold (\ref{l_e}) at an equilibrium point $E^{1}_{\pm},  E^{2}_{\pm}$ is
\begin{align}
T_{E^{1,2}_{\pm}} {\cal E} = &\{(\rho_1, \rho_2, \rho_3, \rho_4)\, |\,\, \cos^3\theta_0 V(\theta_0) \rho_1 - v_0 \cos^3 \theta_0 \rho_2  \nonumber \\
& + [\sin\theta_0 (2v^2_0\cos^2\theta_0 + C^2 -2 U(\theta_0))  \nonumber\\
& + \cos\theta_0 U'(\theta_0) ] \rho_3=0 \}\, . 
\end{align}
Then a basis for$T_{E^{1,2}_{\pm}} {\cal E}$ is given by $\xi_1 = (1, 0, 0, 0)$, $\xi_3 = (0, 0, 1, 0)$ and $\xi_4 = (0, 0, 0, 1).$
A representative of $\bar{J}$ in this basis is
\begin{equation}\label{repres_J_1}
\bar{J}=\left(
\begin{array}{cccc}
\frac{\pm v_0 \cos^2 \theta_0}{\sqrt{U(\theta_0)}} & 0 & 0 \\ \\
 0 & 0 & 1 \\ \\
\frac{V'(\theta_0) \cos^4 \theta_0}{U(\theta_0)}  & a & 0
\end{array}
\right).
\end{equation}The eigenvalues are  
\begin{align*}
\lambda_1 &= \frac{ v_0 \cos^2 \theta_0}{\sqrt{U(\theta_0)}}\,\,\,\,\,\,\,\,\, \text{for}\,\,\, E^1_{+}\,\,\,\text{and}  \,\,\,E^2_{+}\,,\\
\lambda_2 &= \frac{- v_0 \cos^2 \theta_0}{\sqrt{U(\theta_0)}}\,\,\,\,\, \text{for}\,\,\, E^1_{-}\,\,\,\text{and}  \,\,\,E^2_{-}\,.
\end{align*}
and $\lambda_{2,3} = \pm i \sqrt{-a}\,.$

\smallskip

Thus we have proven:

\smallskip

\begin{proposition}\label{prop1}
For every fixed energy level $h$ and any fixed angular momentum 
$|C|\in \left[0, \sqrt{2U(0)} \right)$, the equilibria  $P_{+}$  ($P_-$) have a one-dimensional unstable (stable)  manifold and a  two-dimensional  centre manifold.
\end{proposition}

\smallskip

\begin{proposition}\label{prop2}
For every fixed energy level $h$ and any fixed angular momentum 
$|C|\in \left[0, \sqrt{2U_m} \right)$, the equilibria  $E_{+}^{1,2}$  ($E_{-}^{1,2}$) have a one-dimensional unstable (stable)  manifold and a  two-dimensional  centre manifold.
\end{proposition}

\smallskip

\begin{proposition}\label{prop3}
For every fixed energy level $h$ and any fixed angular momentum 
$|C|> \sqrt{U(0)}$, 
the triple collision manifold is reached (asymptotically) by solutions with double collision as   limit configuration (i.e., the limit configuration has $R=0$).
\end{proposition}

\smallskip

\begin{remark} \label{gamma} 
When $\gamma_0\geq 16 \gamma$, the functions $V(\theta)$ and $W(\theta)$ lose their  critical points at $\theta \neq 0$, and consequently,  the collision manifold does not display a  ``hump''.  The only equilibria on ${\cal C} \setminus {\cal B}_{l,r}$ are those at $P_{\pm}$.
\end{remark}

\subsection{Homographic motions}

\begin{figure}[h]
\centerline
{\includegraphics[scale=0.4]{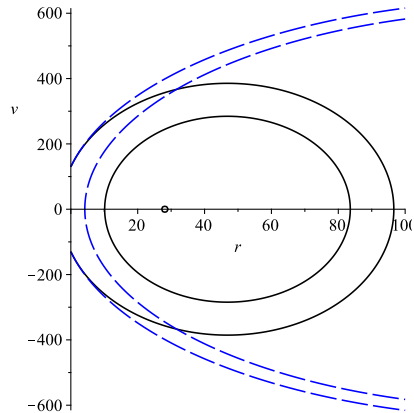}}
{\caption{Homographic motions. Orbits with $h<0$  and $h>0$ are represented with solid lines and dashlines, respectively. 
\label{homo-fig}
}}
\end{figure}

Using similar arguments as in \cite{Arredondo14}, one may prove that motions ejecting/ending from/to the equilibria $P_{\pm}$ are homographic, i.e., they maintain  a self-similar shape of the triangle formed by  the three bodies.  In the Manev isosceles problem, homographic motions form the invariant manifold 
\begin{equation}
{\cal H}:=\{(r,v, \theta,w)\,|\, \theta=0\,, w=0 \}
\label{homo}
\end{equation}
of the system \eqref{sys_MP2}-\eqref{sys_MP2-2}, and the dynamics on ${\cal H}$ are given by
\begin{align}
r^{\prime}&=\frac{\cos^2 \theta}{\sqrt{U(\theta)}}rv,
\label{homo-1}\\
v^{\prime}&= r(2hr + V(\theta)) \frac{\cos^2 \theta}{\sqrt{U(\theta)}}\,.
\label{homo-2}
\end{align}
with  the energy integral
\begin{equation}
v^2+ 2(-h)r^2 -2rV(0) +C^2-2U(0)=0\,.
\label{energy-homo}
\end{equation}
Since on ${\cal H}$ we have $\theta(t)=0$ for all $t$,  physically homographic motions  have a   linear  configurations, with body $m$ positioned midway between the other two.
For $h<0$ we re-write the energy relation \eqref{energy-homo}  as
\begin{align}
\frac{v^2}{2(-h)} &+ \left(   r -\frac{V(0)}{2(-h)} \right)^2  \nonumber \\
&+ \frac{1}{2(-h)}
\left(C^2-2U(0) -\frac{V^2(0)}{2(-h)}\right)=0\,.
\end{align}
and notice that the motion is possible only for momenta $C$ such that
 \begin{equation}
|C|< \sqrt{2U(0) +\frac{V^2(0)}{2(-h)}}\,.
\end{equation}
We also observe that for  
 \begin{equation}
 \sqrt{2U(0)}<|C|<  \sqrt{2U(0) +\frac{V^2(0)}{2(-h)}}
\end{equation}
 all orbits are  periodic and non-collisional, and surround  the equilibrium   located at
\begin{equation}
S=\left( \frac{V(0)}{2(-h)}\,, 0  \right)\,.
\end{equation}
As mentioned, in physical space,  homographic motions correspond to motions with linear configuration. A \textit{homographic equilibrium} is a  (rotating) steady state with the outer bodies rotating at a fixed distance from the central body.   \textit{Homographic periodic orbits} are motions in which  the outer bodies rotate and   ``pulsate" between a maximum and minimum distance from the central body.
For $h>0$ all homographic orbits are unbounded. They either eject/fall into the  collision manifold or  come from infinity, attain a configuration minimal size,  and  return to infinity.  A sketch of the phase portrait of homographic motions is given in Figure \ref{homo-fig}.
 
\subsection{Other aspects of the global flow}
 
 \begin{proposition}
For every fixed $h<0$  and \newline $|C|\in \left(\sqrt{2U_m}, \sqrt{2U(0)}   \right)$ the set 
${\cal C} \setminus \left({\cal B}_{l,r}\cup  {P_{\pm}} \right)$
is not an attractor.
 \end{proposition}
 
 \smallskip
 \noindent \textbf{Proof:} Let $h<0$  and $|C|\in \left(\sqrt{2U_m}, \sqrt{2U(0)} \right)$ be fixed.   In this case the collision manifold and its flow are depicted in Figure \ref{CM_C_in_Um_UM}. The evolution  of the 
 $r$ and $v$ variables is driven by the equations \eqref{sys_MP2} and \eqref{sys_MP2-1}; for reader's convenience we re-write these equations below
\begin{align}
r^{\prime}&=\frac{\cos^2 \theta}{\sqrt{U(\theta)}}rv, \label{equa-1} \\
v^{\prime}&= 2h \left(\frac{\cos^2 \theta}{\sqrt{U(\theta)}} \right)r^2   +2 V(\theta)) \left(\frac{\cos^2 \theta}{\sqrt{U(\theta)}} \right)\,. \label{equa-2}%
\end{align}
We will show that for the given $h$ and $C$ no orbit can tend to ${\cal C} \setminus \left({\cal B}_{l,r}\cup  {P_{\pm}} \right).$
Assume that there is an orbit that  approaches asymptotically ${\cal C} \setminus \left({\cal B}_{l,r}\cup  {P_{\pm}} \right).$   This means that from some $t_0$ the function $r(t)$ is monotone decreasing for all $t > t_0.$ Looking at  \eqref{equa-1}, this implies that $v(t)<0$ for all $t > t_0$. Since $h$ is finite,   the term $\displaystyle{\frac{\cos^2 \theta}{\sqrt{U(\theta)}}}$ bounded and $V(\theta)>0$ for all $\theta,$ for $r$ small enough the right hand side of the \eqref{equa-2} becomes positive, so making $v'>0$. Then   $v$ starts  increasing, 
 becoming  positive again for some $t_1>t_0$, and thus implying that  $r$ is increasing for $t>t_1$. But this contradicts the assumption that $r (t)$ is decreasing for all $t>t_0.$ 
 $\square$

 \smallskip
 
\begin{corollary}
For every fixed $h<0$  and \newline $|C|\in \left(\sqrt{2U_m}, \sqrt{2U(0)}   \right)$, the triple collision manifold is reached (asymptotically)  by solutions for which the limit configuration have zero area, i.e., by solutions with limit configurations that are either linear ($Z=0$), or vertical, with the equal mass bodies in double collision ($R=0$).
\end{corollary}

{
 \smallskip
\begin{corollary} 
For every fixed $h<0$  and \newline $|C|\in \left(\sqrt{2U_m}, \sqrt{2U(0)}   \right)$,   there are no triple collision orbits trapped asymptotically in a near-collision quasi-periodic behaviour. Specifically, there are no orbits for which the outer mass points rotate with increasing spin about the centre of mass, whereas the middle mass point oscillates up and down, with the triple collision being attained asymptotically in time.
\end{corollary}
}
\smallskip
Using Propositions \ref{prop1} and \ref{prop2}  we also deduce

\smallskip
\begin{proposition}
For any $h<0$ fixed and low angular momenta $|C|< \sqrt{2U_m}$ the triple collision  is attainable (either as a ejection or collision) by solutions with spatial and linear limit configurations. 
\end{proposition}

\smallskip

A direct analysis of the  system \eqref{sys_MP2}  also implies that

\smallskip

\begin{proposition} For $h>0$, all orbits are unbounded.
\end{proposition}

{
\section{Conclusions}

   As mentioned in the Introduction,  physical reasoning suggests that  in the $N$-body problem with a binary interaction law of the form $-1/r-B/r^{\alpha}$, $\alpha>1$, $B>0$,  the Manev potential $\alpha=2$ marks a  threshold between two distinct types of near-collision dynamics.
    This was previously proven for $N=2$; see \cite{Diacu00, Stoica00} and the references within. 
     The results of the present paper asserts the same for the isosceles three-body case.   

\begin{enumerate}

\item In the Newtonian case $\displaystyle{U(r)=-1/r}$ \cite{Devaney80}, the total collision is attained only for zero angular momentum. The collision manifold is a sphere with four points removed and its the flow on  is gradient-like with respect to the variable $v$, that is the rate of change of a variable measuring  the size  of triangle formed by the mass points. Double collisions are regularizable and all orbits leading to triple collision are homothetic (self-similar and non-rotating). 

We conjecture this picture does not change if 
 the potential is modified to be to $\displaystyle{U(r)=-1/r^{\alpha}}$ or $\displaystyle{U(r)=-1/r + B/r^{\alpha}}$ with $<1<\alpha<2$ and $B>0$.

 \smallskip 
 \item In the Manev case $\displaystyle{U(r)=-1/r + B/r^{2}}$,  $B>0$, we have that:

 \begin{enumerate}
\item  for momenta size $0\leq C<C_1$, where $C_1 \neq 0$ is a   constant depending on the masses and the gravitational coefficients, the collision manifold has the same topology as in the Newtonian case, but $v$  is constant along any orbit. This was observed by \cite{Diacu93} for $C=0.$ In the present work  we showed that this is the case   for momenta $0<C<C_1$  and we also noted that double collisions are not regularizable;

\item for momenta size $C_1<C<C_2$,  where $C_2\neq 0$ is a  constant depending on the masses and the gravitational coefficients, the collision manifold is formed by the union of a manifold, ${\mathcal C}$, topologically equivalent to a sphere with two (double-collision) lines ${\mathcal{B}}_{l,r}$. On ${\mathcal C}$,  $v$ is constant along any orbit. The lines  ${\mathcal{B}}_{l,r}$ are formed by fixed points. Total collision is attainable  by homothetic and homographic orbits and the (outer) equal masses display black-hole-type motion.

\item  for momenta size $C= C_2$ the collision manifold ${\mathcal C}$ consist in the union of a point and the double collision lines  ${\mathcal{B}}_{l,r}$ of fixed points;

\item  for momenta $C> C_2$ the manifold ${\mathcal C}$ is formed by the double collision lines ${\mathcal{B}}_{l,r}$ of fixed points;

 We conjecture that for $C\neq 0$, there also exist 
 triple collision orbits that are not homographic, similar to the Schwarzschild case discussed below;

\end{enumerate}

\smallskip
\item In the Schwarzschild case $\displaystyle{U(r)=-1/r + B/r^{3}}$, $B>0$  \cite{Arredondo14}, the  total collision is attainable by trajectories  of all angular momenta. The collision manifold is a topologically a sphere with four points removed, similar to  the zero momentum in Newtonian case and low momenta in the Manev case. The flow is gradient-like in $v$, but going in opposite direction with respect to the Newtonian case, and binary collisions are not regularizable. For $C\neq 0$, there exist  triple collision orbits that are both homographic and not.

\end{enumerate}

}

\section*{Acknowledgments}CS was supported by an NSERC Discovery Grant.
%

\end{document}